\documentclass[aps,prb,twocolumn,superscriptaddress,longbibliography]{revtex4-2}
\usepackage{graphicx}
\usepackage{amssymb,amsmath}
\usepackage{bm}
\usepackage{dcolumn}
\usepackage{subfigure}
\usepackage{float}
\usepackage[OT1]{fontenc} 
\usepackage{url}
\usepackage{mathrsfs}
\usepackage{slashed}
\usepackage{color}
\usepackage{verbatim}
\usepackage{enumitem}
\usepackage{txfonts}
\usepackage{soul,physics}
\usepackage[driverfallback=dvipdfm]{hyperref}
\hypersetup{pdfpagemode=FullScreen,colorlinks=true,breaklinks,urlcolor=blue,linkcolor=blue,citecolor=blue}
\usepackage{natbib}
\usepackage{environ}

\usepackage{subfigure}
\usepackage{comment}
\usepackage{hyperref}
\usepackage{amssymb}
\usepackage{bm}
\usepackage{graphicx}
\usepackage{amsmath,amssymb}
\usepackage{tikz,fp}
\usepackage{tikz-cd}
\usetikzlibrary{arrows}
\usetikzlibrary{intersections}
\usetikzlibrary{shapes.geometric}
\usetikzlibrary{decorations.pathmorphing, patterns,shapes,fixedpointarithmetic}
\usetikzlibrary{decorations.markings}

\renewcommand{\vec}[1]{{\bm #1}}

\newcommand{\bea}{\begin{equation}\begin{aligned}}
\newcommand{\eea}{\end{aligned}\end{equation}}

\begin{document}
  
  \title{  Scrambling Transition in Free Fermion Systems Induced by a Single Impurity}
  
 \author{Qucheng Gao}
\affiliation{Department of Physics, Boston College, Chestnut Hill, Massachusetts 02467, USA}
 \author{Tianci Zhou}
\affiliation{Department of Physics, Virginia Tech, Blacksburg, Virginia 24061, USA}
\author{Pengfei Zhang}
\email{PengfeiZhang.physics@gmail.com }
\affiliation{Department of Physics, Fudan University, Shanghai, 200438, China}
\affiliation{Shanghai Qi Zhi Institute, AI Tower, Xuhui District, Shanghai 200232, China}
\author{Xiao Chen}
\email{chenaad@bc.edu}
\affiliation{Department of Physics, Boston College, Chestnut Hill, MA 02467, USA}

  \begin{abstract}
  In quantum many-body systems, interactions play a crucial role in the emergence of information scrambling. When particles interact throughout the system, the entanglement between them can lead to a rapid and chaotic spreading of quantum information, typically probed by the growth in operator size in the Heisenberg picture. In this study, we explore whether the operator undergoes scrambling when particles interact solely through a single impurity in generic spatial dimensions, focusing on fermion systems with spatial and temporal random hoppings. By connecting the dynamics of the operator to the symmetric exclusion process with a source term, we demonstrate the presence of an escape-to-scrambling transition when tuning the interaction strength for fermions in three dimensions. As a comparison, systems in lower dimensions are proven to scramble at arbitrarily weak interactions unless the hopping becomes sufficiently long-ranged. Our predictions are validated using both a Brownian circuit with a single Majorana fermion per site and a solvable Brownian SYK model with a large local Hilbert space dimension. This suggests the universality of the theoretical picture for free fermion systems with spatial and temporal randomness.

  \end{abstract}
  
  \maketitle

  \tableofcontents

  \section{Introduction}
  Generic interacting many-body systems can serve as their own bath, a pivotal element for the manifestation of quantum thermalization in isolated systems \cite{PhysRevA.43.2046,PhysRevE.50.888}. This involves the obscuring of all local initial conditions within the entire system after prolonged evolution, measured by the growth in operator size \cite{Sekino_2008,Hayden_2007,Shenker:2013pqa,Shenker:2014cwa,Roberts:2014isa,Maldacena:2015waa,Roberts:2018mnp}. In this process, interactions play a crucial role. In the absence of interactions, excitations with infinite lifetimes can carry quantum information, remaining free from dissipation \cite{Calabrese_2005}. This occurs because quadratic Hamiltonians conserve the number of field operators, thereby preventing the growth of operator size. However, in the presence of interactions, a single excitation can undergo scattering, giving rise to multiple excitations. This iterative process leads to a rapid increase in complexity for simple initial operators subject to the Heisenberg evolution \cite{Nahum:2017yvy,Hunter-Jones:2018otn,vonKeyserlingk:2017dyr,Khemani:2017nda,Qi:2018bje,Dias:2021ncd,PhysRevResearch.3.L032057,Qi:2019rpi,Zhou:2021syv,Omanakuttan:2022ikz,Zhang:2022fma,Liu:2023lyu,PhysRevLett.122.216601,Lucas:2020pgj,Chen:2020bmq, Chen:2019klo,Yin:2020oze,Xu_2019,Zhou_2019,Chen_2019,Zhou_2020,lashkari2013towards}.

  Recently, new insights into this problem have been gained from the study of fermionic systems that only interact through a single impurity \cite{gao2023information}. The key question is whether a single impurity can effectively scramble the entire system. While one might typically anticipate that introducing a local impurity into a large many-body system does not lead to significant changes in bulk dynamics, it becomes apparent that even with just a single interaction, the Hamiltonian is no longer quadratic, thus permitting the growth of operator size. Indeed, previous study unveils the  emergence of information scrambling in 1D systems with spatial and temporal random short-range hoppings for arbitrary weak interaction strength at the impurity \cite{gao2023information}. (Many other papers have investigated similar setups---non-equilibrium dynamics in the presence of boundary perturbations. See Ref. \cite{blythe2007nonequilibrium} for a review on classical stochastic processes and see Ref. \cite{landi2022nonequilibrium} for a review on quantum systems. Here, we study this from the perspective of quantum information. We also note a paper that studies information scrambling in an integrable Kondo model \cite{dora2017information}.) In this setup, both the operator and entanglement exhibit diffusive scaling, stemming from the random walk characteristics of fermion operators on the 1D lattice. However, it is acknowledged that the properties of random walks are greatly influenced by the connectivity among different sites across various geometries.
    
  \begin{figure}
    \centering
    \includegraphics[width=0.98\linewidth]{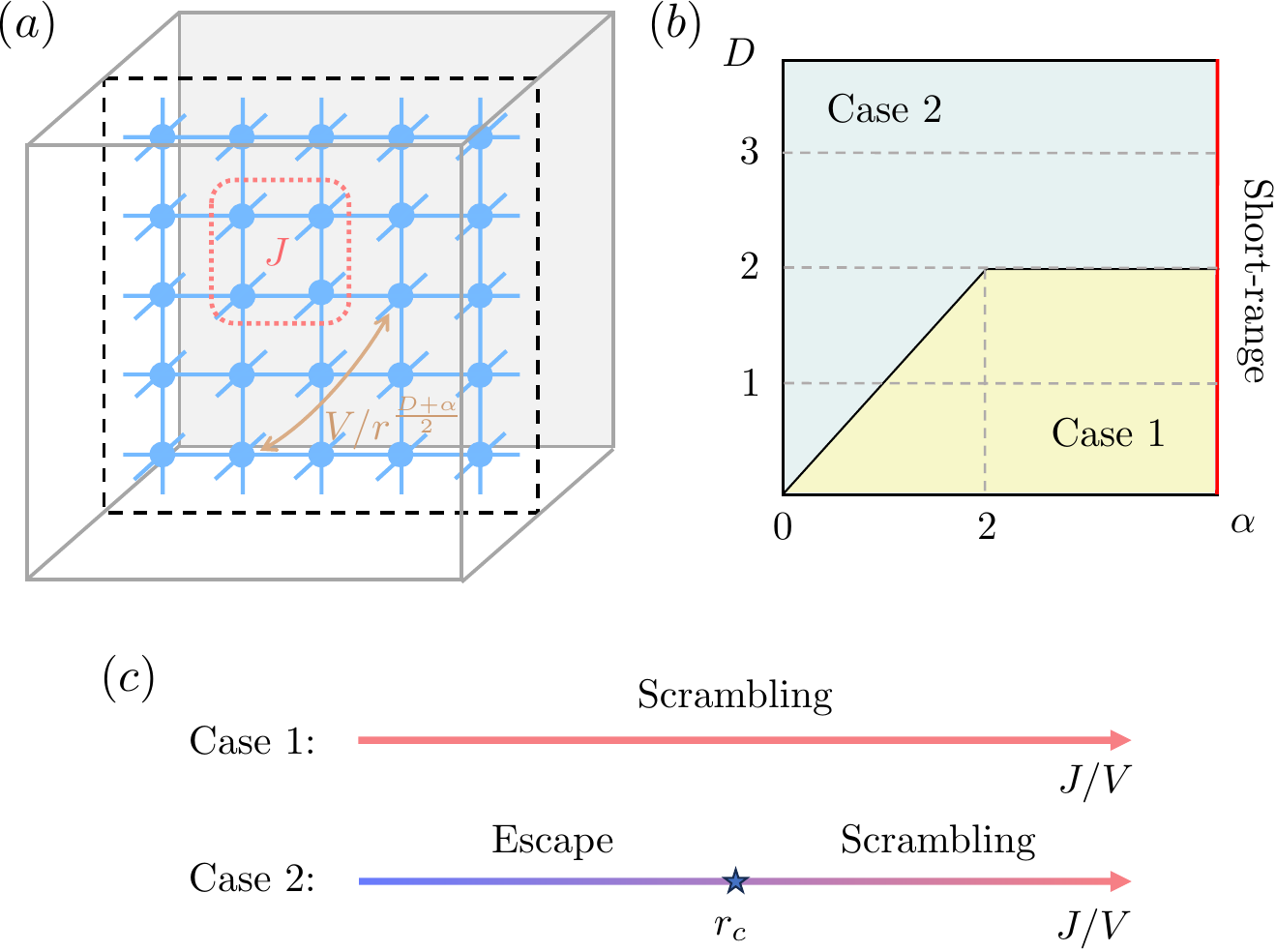}
    \caption{(a). The schematics of our model in $D=3$ with a single Majorana fermion per site. The model includes a solitary random interaction with strength $J$ and Brownian hopping that decays as $1/r^{\frac{D+\alpha}{2}}$ for $\alpha > 0$. The case of short-range hopping corresponds to $\alpha \rightarrow \infty$. [(b) and (c)]. The phase diagram of the model. Here, we extend the dimension $D$ to an arbitrary real number. In the case 1 regime, arbitrarily weak interaction can scramble the entire system, whereas in the case 2 regime, a critical interaction strength is necessary to realize the scrambling of operators near the interacting sites.
 } \label{fig1}
  \end{figure}

  Building upon this insight, we delve deeper into investigating the growth of operator size within such systems across arbitrary dimensions. Utilizing a phenomenological percolation model on trees, we identify a notable distinction between systems in 3D and lower dimensions, summarized in Fig. \ref{fig1}. In 3D, interactions do not lead to a persistent growth of operator size unless their strength exceeds a critical value. Within this non-scrambling regime, the operator quickly escapes from the impurity, following the P\'olya's theorem, and we term this as the escape phase. For stronger interactions, the operators near interacting sites can scramble into non-local operators, analogous to the scenario in lower dimensions with short-range hoppings. Additionally, when the hopping range is long enough, we also identify similar dynamical transitions in lower dimensions through the utilization of L\'evy flight properties. Our predictions are demonstrated by both numerical simulations based on a small $N$ Brownian circuits \cite{Xu_2019,Chen_2019,Zhou_2019,Zhou_2020,lashkari2013towards} and analytical calculations in a solvable large $N$ Brownian Sachdev-Ye-Kitaev (SYK) chain \cite{SY,kitaev2015simple,Maldacena:2016hyu,Kitaev:2017awl,Gu_2017,Saad:2018bqo,Sunderhauf:2019djv,Chowdhury:2021qpy}. This suggests universality regardless of the dimensionality of the local Hilbert space. 

  \section{Brownian circuits}
  To be concrete, we first examine Brownian circuits with nearest-neighbor hoppings in a generic dimension $D$, where operator dynamics can be mapped to a classical stochastic process. As we will demonstrate, the physical picture obtained in this model is also applicable to Brownian SYK models with large Hilbert space dimensions. Our focus is on Brownian circuits of Majorana fermions. Each site hosts a single Majorana mode $\chi_{\bm{x}}$ with canonical anti-commutation relations $\{\chi_{\bm{x}},\chi_{\bm{y}}\}=2\delta_{\bm{x}\bm{y}}$. The Hamiltonian reads
  \begin{equation}\label{eq:Hcircuit}
  dH(t) = i\sum_{\langle \bm{x} \bm{y}\rangle} dV_{\bm{x},\bm{y}}~\chi_{\bm{x}}\chi_{\bm{y}}+dJ \prod_{\bm{x}\in \square} \chi_{\bm{x}},
  \end{equation}
  where the first term denotes the 
hopping of the free fermion between neighboring sites and the second term is the interaction term. Here $\square$ labels four sites in a single plaquette near the origin, as illustrated in Fig. \ref{fig1} with $D\geq 2$. For $D=1$, we can simply pick four contiguous sites near the origin. Independent Brownian variables $dV_{\bm{x},\bm{y}}$ and $dJ$ satisfy the Wiener process, with
  \begin{equation}
    \overline{dV_{\bm{x},\bm{y}}dV_{\bm{x}',\bm{y}'}}= Vdt ~\delta_{\bm{x}\bm{x}'}\delta_{\bm{y}\bm{y}'}, \ \ \ \ \ \ \overline{dJ^2}= Jdt.
  \end{equation}
  In a short time interval $dt$, the unitary evolution is given by $dU=e^{-i dH}$. We are interested in the operator dynamics, governed by the Heisenberg evolution $O(t+dt)=dU^\dagger O(t)dU$. To study the growth of operator size, we introduce a complete orthonormal basis of Hermitian operators $\{B_\mu\}=\{i^{q(q-1)/2}\chi_{\bm{x}_1}\chi_{\bm{x}_2}...\chi_{\bm{x}_q}\}$. Each Majorana string $B_\mu$ can be labeled by its height $\mathbf{h}_\mu$, defined as $h_{\mu,\bm{x}}=1$ for $\bm{x}\in\{\bm{x}_1,\bm{x}_2,...,\bm{x}_q\}$ and otherwise $h_{\mu,\bm{x}}=0$. The size of $B_\mu$ is further defined as 
  $n_{\mu} = \sum_{\vec{x} } h_{\mu, \vec{x}}$

 We expand $O(t)$ in this set of basis operators as $O(t)=\sum_\mu \alpha_\mu(t) B_\mu$, where $\alpha_\mu(t)$ represents the wave function for the operator evolution. In Brownian circuits, the phase of $\alpha_\mu(t)$ is averaged out due to the temporal randomness, and the evolution can be formulated in terms of a classical stochastic process described by the probability distribution $f_\mu(t)=|\alpha_\mu(t)|^2$, which is normalized $\sum_\mu f_\mu(t)=1$ due to unitarity \cite{Roberts:2014isa,Roberts:2018mnp}. The size of $O(t)$ is defined as $N(t)\equiv\sum_\mu n_\mu f_\mu(t)$. By generalizing the analysis in Ref.~\cite{gao2023information}, this dynamics describes a symmetric exclusion process (SEP) with a single source term at the origin~\cite{liggett1985interacting,spitzer1991interaction}. The governing master equation for this dynamics, along with its derivation details, are provided in the Appendix. Based on this master equation, we conduct numerical simulation using the following update rules: 
 \begin{enumerate}
  \item We implement an unbiased random walk for each particle independently. 

  \item Each time when one particle returns to the origin, we branch it by adding $n_i$ particles with a probability $p_i$.

  \end{enumerate}
  In the original model \eqref{eq:Hcircuit}, we expect $n_i=2$ and $p_i\propto J/V$. The system is initialized by putting a single particle at the origin, and the operator size growth is studied by counting the number of particles. 

  \begin{figure}[t]
    \centering
    \includegraphics[width=0.9\linewidth]{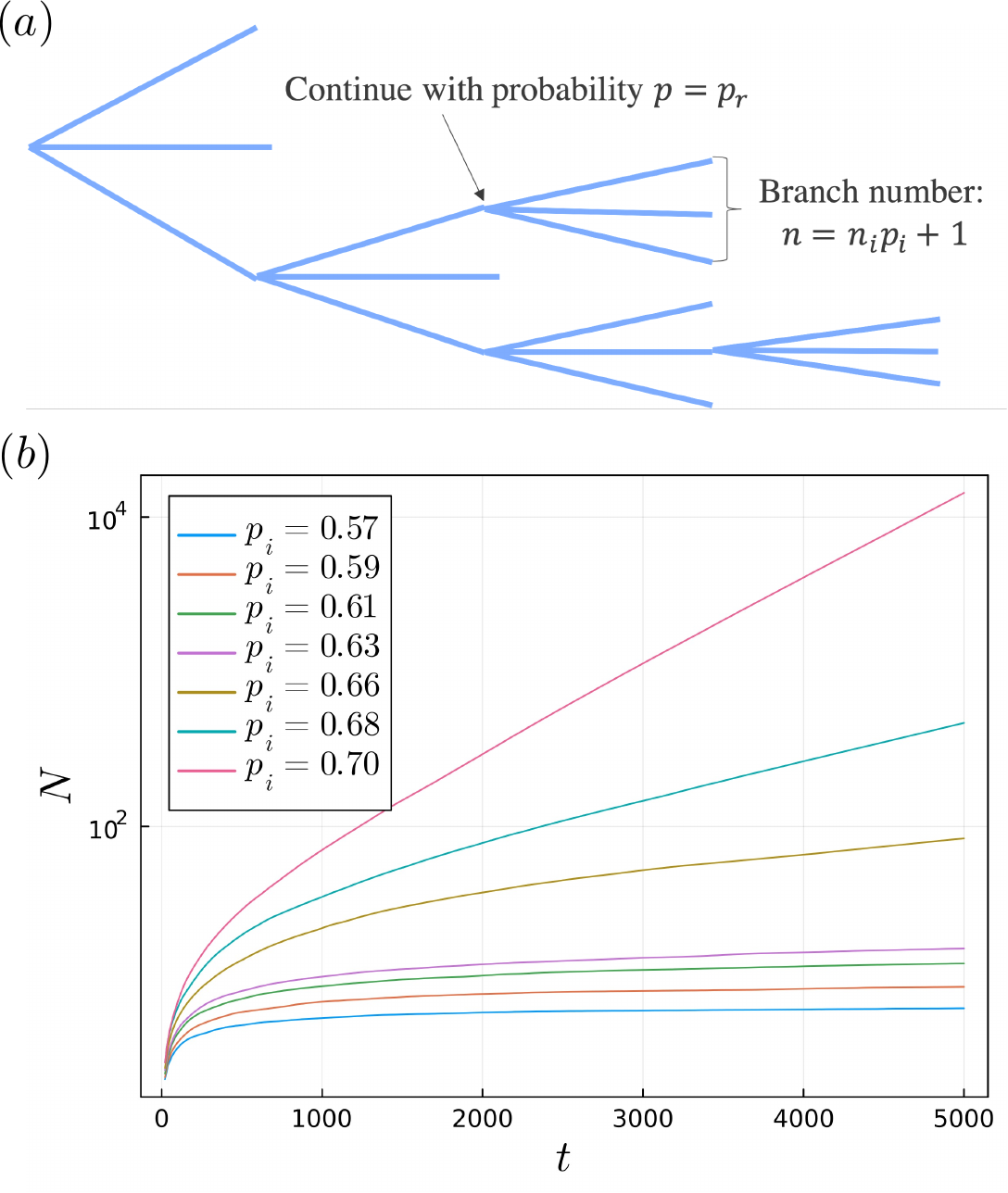}
    \caption{(a). Schematics of the effective percolation model on trees: each edge represents a particle at the origin, which is removed with a probability of $1-p_r$ if it escapes. Each remaining edge branches into $n=n_i p_i+1$ edges. (b). Particle number $N$ as a function of time $t$ in log-lin scale in 3D. We average over $10000$ samples. We observe a transition of particle number between $p_i = 0.63$ and $p_i = 0.66$. The theoretical prediction of $p_i^c$ is $0.646$.} \label{fig2}
  \end{figure}

  We note that in 1D and 2D, an operator originating from the origin exhibits diffusive spreading, independent of the parameter $p_i$, eventually encompassing the entire space. Complexity arises in 3D, where for large $p_i$, the operator continues to grow over time. However, decreasing $p$ can mitigate its growth. Notably, for $p<p_c$, during time evolution, the operator's size saturates to a finite constant, indicating the presence of a scrambling transition as $p_i$ varies. 

  To understand the presence or absence of the transitions across various dimensions, we recognize that the dynamics can be modeled as percolation on a tree. As shown in Fig. \ref{fig2} (a), in this model, each vertex is branching into $n$ vertex, with each edge subject to removal with probability $1-p$. It is known that this model exhibits a percolation transition at $pn=1$~\cite{lyons1990random}. When $pn>1$, the root is connected to an infinite number of the vertices in the tree. Conversely, if $pn<1$, the root connects to only a finite number of vertices in the tree. Our model maps to the percolation model as follows: each edge represents a particle from the origin, with each vertex's branching number estimated as $n=n_ip_i+1$. Edges from a vertex are removed together with probability $1-p_r$ if the particle doesn't return to the origin. Here $p_r$ represents the return probability of a random walker to return to its starting point. The transition of this model is expected to occur at $p_rn=1$. 
  
  Given $n=n_ip_i+1\geq 1$, observing a scrambling transition requires $p_r$ to be less than 1. In 1D and 2D, where the random walker returns to its starting point with probability one, there is no scrambling transition, and the model remains in the scrambling phase as long as $p_i>0$.  On the other hand,  according to P\'olya's theorem~\cite{polya_uber_1921,polya},
  the 3D random walker will return to its starting point with probability $p_{r} \approx 0.341$. A brief review of the derivation is presented in the supplementary material for clarity. The probability of returning being less than one allows us to induce a scrambling transition at some finite $p_i^c$, such that  $(n_ip_i^c+1)p_r=1$. For $p_i<p_i^c$, particles escape from the origin, leading to saturation of the operator size. Specifically, we find $p_i^c\approx 0.646$ for $n_i=3$, which aligns with the results from 3D classical particle model simulations illustrated in Fig. \ref{fig2} (b). 
  
  We can generalize the above discussion to fermion operators that are initially not located at the origin. As demonstrated in the supplementary material, the probability of reaching the origin is inversely proportional to the distance $r$ from the origin. This suggests that these operators have a reduced probability of reaching the origin. Nevertheless, upon returning to the origin, they may undergo scrambling if $p_i > p_i^c$.

  \section{Brownian SYK model}
  We now develop an analytically solvable large-N model capable of demonstrating a scrambling transition. Specifically, we consider the Brownian SYK model \cite{Saad:2018bqo,Sunderhauf:2019djv} with a single interacting impurity. In this model, we have $N$ Majorana fermions $\chi_{\bm{x},i}$ with $i\in\{1,2,...,N\}$ on each site, which satisfies $\{\chi_{\bm{x},i},\chi_{\bm{y},j}\}=2\delta_{\bm{x}\bm{y}}\delta_{ij}$. The Hamiltonian reads
  \begin{equation}
  \begin{aligned}
  dH(t) = &i\sum_{\langle \bm{x} \bm{y}\rangle,ij} dV_{\bm{x},\bm{y}}^{ij}~\chi_{\bm{x},i}\chi_{\bm{y},j}\\&+i^{q/2}
  \sum_{i_1\leq i_2\leq...\leq i_q }dJ_{i_1i_2...i_q} \chi_{\mathbf{0},i_1}\chi_{\mathbf{0},i_2}...\chi_{\mathbf{0},i_q},
  \end{aligned}
  \end{equation}
  where the second term represents a $q$-body interaction on the impurity with $q\geq 4$. Brownian variables with different indices are independent and satisfy
  \begin{equation}\label{eqn:Brownian}
  \overline{(dV_{\bm{x},\bm{y}}^{ij})^2}={V dt}/{4N},\ \ \ \ \ \ \overline{(dJ_{i_1i_2...i_q})^2}={(q-1)!J dt}/{4N^{q-1}}.
  \end{equation}
  The model can be analyzed using the large-$N$ expansion. Focusing on its real-time dynamics, we first introduce the retarded Green's functions $G^R_{
  \bm{x}}(t)\equiv -i\theta(t)\langle \{\chi_{\bm{x},i}(t),\chi_{\bm{x},i}(0)\}\rangle$, where $\theta(t)$ is the Heaviside step function. In SYK-like models, the self-energy is dominated by melon diagrams, which gives $\Sigma^R_{\bm{x}}(\omega)=i(z V+J\delta_{\bm{x},\bm{0}})/4\equiv i\Gamma_{\bm{x}}/4$. Here, $z=2D$ is the coordination number for the square lattice and $\Gamma_{\bm{x}}$ represents the quasi-particle lifetime on site $\bm{x}$. Transforming into the time domain, we find $G^R_{\bm{x}}(t)=-2ie^{-\Gamma t/2}$.
  
  In large-$N$ systems, due to the large on-site Hilbert space, the operator size usually experiences exponential growth in the early-time regime, characterized by a rate known as the quantum Lyapunov exponent $\varkappa$. Consequently, we are examining whether the system undergoes scrambling by assessing the existence of a non-vanishing $\varkappa$. It is known that the average size is related to out-of-time-order (OTO) commutator as \cite{Roberts:2018mnp,Qi:2018bje}
  \begin{equation}
  N_{\bm{x}}(t)=\frac{1}{4}\sum_{j} \langle |\{\chi_{\bm{y=0},i}(t),\chi_{\bm{x},j}(0)\}|^2\rangle,
  \end{equation}
   where we introduce a subscript $\bm{x}$ to denote the location of the operator. In SYK-like models, the self-consistent equation for the OTO commutator arises from the ladder diagram \cite{Maldacena:2016hyu,Zhang:2020jhn}
  \begin{equation}\begin{aligned}
  N_{\bm{x}}(t_1)=&-\int dt'~G^R_{\bm{x}}(t_{12})^2 \\
  &\times\left[\frac{V}{4}\sum_{\langle \bm{x}\bm{y}\rangle}N_{\bm{y}}(t_2)+\frac{J(q-2)}{4}\delta_{\bm{x},\bm{0}} N_{\bm{x}}(t_2)\right].
\end{aligned}
  \end{equation}
Here we neglect the inhomogeneous terms since it does not contribute to the asymptotic behavior. In Brownian models, this is equivalently expressed as a differential equation:
  \begin{equation}\label{eq:syk_diff}
  -\frac{ dN_{\bm{x}}}{dt}=V\left(zN_{\bm{x}}-\sum_{\langle \bm{x}\bm{y}\rangle}N_{\bm{y}}\right)-J(q-2) \delta_{\bm{x},\bm{0}} N_{\bm{x}}.
  \end{equation}
   The first term describes the diffusive spreading, while the second term is a source term at $x=0$, which can potentially result in an exponential growth of $N_x$.
  This equation takes the form of the imaginary-time Schr\"odinger equation on the lattice with an attractive delta potential with depth $\sim J$ at the origin. Assuming $N_{\bm{x}}(t)=\exp(\varkappa t) N_{\bm{x}}$ for sufficiently long time $t$, we recognize that a positive Lyapunov exponent $\varkappa$ corresponds to a bound state with energy $-\varkappa$. The solution is given by the Lippmann-Schwinger equation
  \begin{equation}\label{eqn:Lippmann}
   \frac{V}{J(q-2)}=\int \frac{d^Dk}{(2\pi)^D}\frac{1}{\varkappa/V+2D-\sum_\alpha2\cos (k_\alpha)}.
  \end{equation}
  Here, $\alpha$ labels different spatial directions $\alpha\in\{x,y,...\}$. 
   
  \begin{figure}
    \centering
    \includegraphics[width=0.78\linewidth]{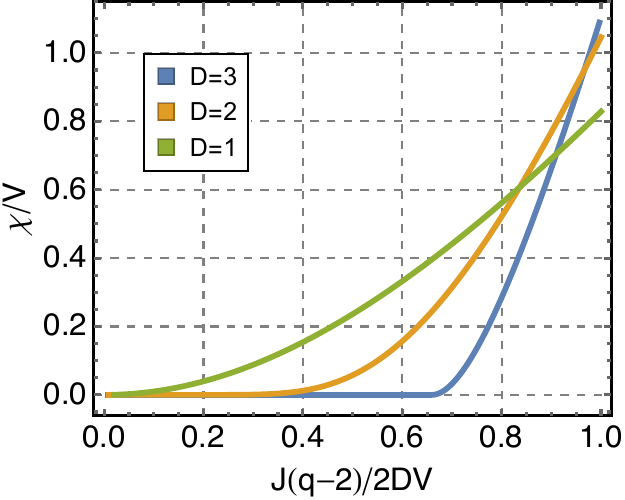}
    \caption{Results for the quantum Lyapunov exponent $\varkappa$ as a function of interaction strength $J/V$ in different dimensions $D\in \{1,2,3\}$, obtained by solving Eq. \eqref{eqn:Lippmann}. The result is consistent with the theoretical prediction that a scrambling transition occurs at the critical strength given by $J(q-2)/2DV=(1-p_r)=0.659$.
 } \label{fig3}
  \end{figure}
  We focus on the regime with a shallow bound state $\varkappa/V\ll 1$, where we can transition to the continuum limit by expanding $1-\cos k\approx k^2/2$. In this limit, Eq. \eqref{eqn:Lippmann} aligns with the standard Schrödinger equation with a quadratic dispersion. It is well-established that in 3D, a finite depth of the potential is required to sustain a bound state. In contrast, in 1D or 2D, a bound state emerges under infinitely weak attractions, leading to exponential growth in operator size. We can further make a direct connection between the bound state problem and the random walk picture in the last section. To determine the critical point $J_*$, we set $\varkappa=0$ on the R.H.S. of \eqref{eqn:Lippmann}. Then, as reviewed in the supplementary material, it can be related to the returning probability $p_r$ by $\text{R.H.S.}=[2(1-p_r)D]^{-1}$. In particular, $p_r=1$ for $D\leq 2$ originates from the divergence of the integral $\sim\int k^{D-1}dk /k^2$. The critical point $J_*$ is then given by $J_*/V=2D(1-p_r)/(q-2)$, a close analog of results in Brownian circuits. For $D=3$, this predicts $J(q-2)/2DV=(1-p_r)=0.659$. We can further obtain the quantum Lyapunov exponent by solving Eq. \eqref{eqn:Lippmann} exactly, using the analytical results for lattice Green's functions \cite{2010JPhA...43D5205G}. The result is plotted in Fig.~\ref{fig3} for $D\in \{1,2,3\}$. In 3D, we can clearly observe a transition from escape phase with $\chi=0$ to scrambling phase with $\chi>0$.

  \section{Long-range hopping}
  We now generalize our results to systems with long-range random hoppings \cite{Chen_2019,Zhou_2020,Zhang:2021mdc,Sahu:2021lgw,Zhang:2021mdc,Sahu:2021lgw,PhysRevLett.128.010604,PhysRevLett.128.010605,PhysRevLett.128.010603}. 
  Since we have demonstrated that both $N=1$ Brownian circuits and the large $N$ Brownian SYK model share the same phase diagram, here we take the Brownian SYK model as an example. The hopping term in the Hamiltonian now becomes
  \begin{equation}
  dH_0(t) = \sum_{ \bm{x}\neq \bm{y},ij} \frac{i}{2|\bm{x}-\bm{y}|^\frac{\alpha+D}{2}}dV_{\bm{x},\bm{y}}^{ij}~\chi_{\bm{x},i}\chi_{\bm{y},j},
  \end{equation}
  which $dV_{\bm{x},\bm{y}}^{ij}$ still has variance \eqref{eqn:Brownian}. Carrying out the similar calculation as in the last section, we find firstly the decay rate becomes $\Gamma_{\bm{x}}=\sum_{\bm{y}\neq\bm{x}}\frac{V}{|\bm{x}-\bm{y}|^{\alpha+D}}+J\delta_{\bm{x},\bm{0}}$. Therefore, to ensure the convergence of the decay rate, we focus on $\alpha>0$. Otherwise, the model is effectively all-to-all connected, whose scrambling dynamics in the presence of a singular interaction has been analyzed in Ref.~\cite{gao2023information}. Secondly, the random walk is replaced by a L\'evy flight. This is reflected in the Lippmann-Schwinger equation, which now leads to 
  $$\frac{V}{J_*(q-2)}=\int \frac{d^Dk}{(2\pi)^D}\frac{1}{\sum_{\bm{y}\neq \bm{0}}|\bm{y}|^{-\alpha-D}[1-\cos ( \bm{k}\cdot \bm{y})]}.$$
  To determine whether the R.H.S. diverges, we first perform the small $k$ expansion of the denominator, which gives
  \begin{equation}
  \begin{aligned}
\sum|\bm{y}|^{-\alpha-D}[1-\cos ( \bm{k}\cdot \bm{y})]=\begin{cases}
                    k^\alpha & \text{if\ \ \ \ }  \alpha < 2,  \\
                     k^2 & \text{if\ \ \ \ } \alpha \geq 2.
                 \end{cases} 
                 \end{aligned}
  \end{equation}
  As a result, the returning probability of the L\'evy flight is $1$ for $D\leq \alpha$ when $\alpha \in (0,2)$ and $D\leq 2$ when $\alpha\in[2,\infty)$. This defines the parameter regime referred to as case 1 in Fig. \ref{fig1}. Conversely, when the integral converges for higher dimensions, an escape-to-scrambling transition typically occurs, denoted as case 2 in Fig. \ref{fig1}. In particular, for $\alpha\geq 2$, the phase diagram is the same as the short-range hopping case. 

  \section{Discussions}
    In this study, we investigate the information dynamics in free fermion systems with spatial and temporal randomness that interacts through a single impurity. Our findings reveal that a solitary interaction can trigger a scrambling phase transition depending on the system's dimensionality and hopping range. We establish a universal phase diagram by employing both $N=1$ Brownian circuits and large $N$ Brownian SYK models. In high dimensions or with a long hopping range, the model undergoes an escape-to-scrambling transition upon tuning the interaction strength $J$. Within the scrambling phase, operators proximate to the interacting site can scramble into an extensive operator under time evolution. On the other hand, in low-dimensional systems with a short hopping range, even arbitrarily weak interactions lead to the scrambling of the entire system. We expect that these problems can also be characterized by other information quantities, such as the entanglement entropy and mutual information ~\cite{gao2023information}.


  The dynamical transitions of information scrambling have been observed in various contexts. Specifically, Refs. \cite{PhysRevLett.130.250401,PhysRevLett.131.220404} identify an environment-induced scrambling transition in systems embedded in environments. One can also treat Majorana fermions in bulk as an environment, achieving the transition by increasing the system-environment coupling $V$. In Refs. \cite{PhysRevLett.130.250401,PhysRevLett.131.220404}, the information never returns to the system once it enters the environment, corresponding to $p_r=0$. Our analysis in this work proposes a refinement of the picture for environments with memory. (A more recent preprint~\cite{gribben2024markovian} studied the scrambing transition where the information backflow can be tuned. Consistent with our results, they didn't find scrambling transition in low dimensions when the information backflow was nonzero, corresponding to $p_r=1$.) The discussions in our paper are also pertinent to the intriguing question of whether a single thermal island can thermalize the entire system, a query crucial for understanding the existence of many-body localization phases. To address this question, it is imperative to extend the current discussions to models with static hopping strength, where localization is feasible. Nevertheless, the theoretical analysis in such cases becomes considerably more challenging, and we defer this task to future works.

\begin{acknowledgements}
    We thank Vikram Ravindranath, Zhenhua Yu and Tian-Gang Zhou for helpful discussions.
    We are especially thankful for the insightful discussions with Paul Krapivsky, who introduced us to their work on non-equilibrium dynamics in the presence of a boundary source term ~\cite{krapivsky2012symmetric,krapivsky2014lattice,bauer2021random,krapivsky2019free,krapivsky2020free}. In particular, in Ref. \cite{bauer2021random}, the classical models they considered closely resemble our current studies. They also observed a phase transition by varying the parameters of the source term. The random walkers in their work correspond to our large $N$ Brownian SYK model, while the SEP corresponds to our $N=1$ Brownian circuits. This research is supported by the National Science Foundation under Grant No. DMR-2219735 (Q. G. and X. C.) and the National Natural Science Foundation of China under Grant No. 12374477 (P. Z.).
\end{acknowledgements}

\bibliography{ref.bib}

\newpage
\onecolumngrid
\pagenumbering{arabic} 
\setcounter{equation}{0}
\setcounter{figure}{0}
\setcounter{table}{0}

\renewcommand{\thepage}{A\arabic{page}}
\renewcommand{\thesection}{A\arabic{section}}
\renewcommand{\theequation}{A\arabic{equation}}
\renewcommand{\thefigure}{A\arabic{figure}}
\renewcommand{\thetable}{A\arabic{table}}

\appendix
\section{Master equation in Brownian circuits}
In this section, we give a detailed derivation of the master equation in Brownian circuits.
The Hamiltonian reads
\bea
    dH(t) = i\sum_{\langle \bm{x} \bm{y}\rangle} dV_{\bm{x},\bm{y}}~\chi_{\bm{x}}\chi_{\bm{y}}+dJ \prod_{\bm{x}\in \square} \chi_{\bm{x}},
\eea
where $dV_{\bm{x},\bm{y}}$ and $dJ$ satisfy the Wiener process, with
\bea
    \overline{dV_{\bm{x},\bm{y}}dV_{\bm{x}',\bm{y}'}}= Vdt ~\delta_{\bm{x}\bm{x}'}\delta_{\bm{y}\bm{y}'}, \ \ \ \ \ \ \overline{dJ^2}= Jdt,
\eea
and $\square$ labels four sites in a single plaquette near the origin as illustrated in the main text.

We can expand the evolution of the operator $O(t)$ to second order:
\bea
    dO(t)  &= e^{idH(t)} O(t) e^{-idH(t)} - O(t)\\
    &=  [ i dH(t), O (t) ]  + \frac{1}{2} [ i dH(t), [ i dH(t), O (t) ] ] \\
    &= i [ dH(t), O(t) ] - \frac{1}{2} \{ dH(t) dH(t), O(t) \}  + d H(t)  O(t) d H (t) \\
    &= i [ dH(t), O(t) ] - \sum_{\langle \bm{x} \bm{y}\rangle} O(t) V dt - O(t) J dt
    -\sum_{\langle \bm{x} \bm{y}\rangle} \chi_{\bm{x}}\chi_{\bm{y}} O(t) \chi_{\bm{x}}\chi_{\bm{y}} V dt + \prod_{\bm{x}\in \square} \chi_{\bm{x}} O(t) \prod_{\bm{x}'\in \square} \chi_{\bm{x}'}Jdt.
\eea
We introduce a complete orthonormal basis of Hermitian operators $\{B_\mu\}=\{i^{q(q-1)/2}\chi_{\bm{x}_1}\chi_{\bm{x}_2}...\chi_{\bm{x}_q}\}$ and the expansion coefficient $\alpha_{\mu}(t)$ is
\bea
    \alpha_{\mu}(t) = \frac{1}{\tr( B_\mu B_\mu ) }\tr( B_{\mu} O(t) ).
\eea
Its time evolution is given by
\bea
    d \alpha_{\mu} (t) &= \frac{1}{\tr( B_\mu^2) }  \tr( B_{\mu} d O(t) ) \\
    &= \frac{i}{\tr( B_\mu^2) }  \tr( B_{\mu} [dH(t), O(t)]   ) 
    - \sum_{\langle \bm{x} \bm{y}\rangle}\alpha_\mu (t) Vdt  - \alpha_\mu (t) Jdt \\
    &- \frac{1}{\tr( B_\mu^2) }\sum_{\langle \bm{x} \bm{y}\rangle}\tr\big( B_{\mu} 
     \chi_{\bm{x}}\chi_{\bm{y}} O(t) \chi_{\bm{x}}\chi_{\bm{y}}\big)Vdt + \frac{1}{\tr( B_\mu^2) }\tr\big( B_{\mu} 
     \prod_{\bm{x}\in \square} \chi_{\bm{x}} O(t) \prod_{\bm{x}'\in \square} \chi_{\bm{x}'}\big)Jdt
     , \\
\eea
here,
\bea
    &\frac{1}{\tr( B_\mu^2) }\sum_{\langle \bm{x} \bm{y}\rangle}\tr\big( B_{\mu} 
     \chi_{\bm{x}}\chi_{\bm{y}} O(t) \chi_{\bm{x}}\chi_{\bm{y}}\big)Vdt \\
     =& \frac{1}{\tr( B_\mu^2) }\sum_{\langle \bm{x} \bm{y}\rangle}\tr\big( \chi_{\bm{x}}\chi_{\bm{y}}B_{\mu} \chi_{\bm{x}}\chi_{\bm{y}} O(t) \big)Vdt \\
     =&-\sum_{\langle \bm{x} \bm{y}\rangle}q_{\mu,\bm{x},\bm{y}}\alpha_\mu (t)Vdt,
\eea
\bea
    &\frac{1}{\tr( B_\mu^2) }\tr\big( B_{\mu} 
     \prod_{\bm{x}\in \square} \chi_{\bm{x}} O(t) \prod_{\bm{x}'\in \square} \chi_{\bm{x}'}\big)Jdt\\
     =&\frac{1}{\tr( B_\mu^2) }\tr\big( \prod_{\bm{x}'\in \square} \chi_{\bm{x}'}
     B_{\mu} \prod_{\bm{x}\in \square} \chi_{\bm{x}} O(t) \big) Jdt\\
     =&q_{\mu,\prod_{\bm{x}\in \square} \chi_{\bm{x}} }\alpha_\mu (t)Jdt, 
\eea
where $q_{\mu,\bm{x},\bm{y}} = 1$ if $\chi_{\bm{x}}, \chi_{\bm{y}}\in B_{\mu}$ or $\chi_{\bm{x}}, \chi_{\bm{y}}\notin B_{\mu}$; $q_{\mu,\bm{x},\bm{y}} = -1$ if $\chi_{\bm{x}}\in B_{\mu}, \chi_{\bm{y}}\notin B_{\mu}$ or $\chi_{\bm{y}}\in B_{\mu}, \chi_{\bm{x}}\notin B_{\mu}$, and $q_{\mu,\prod_{\bm{x}\in \square}\chi_{\bm{x}}} = 1$ if $|~\chi\in \prod_{\bm{x}\in \square} \chi_{\bm{x}} \mid \chi\in B_{\mu}~| \equiv 0 \mod 2$;
$q_{\mu,\prod_{\bm{x}\in \square}\chi_{\bm{x}}} = -1$ if $|~\chi\in \prod_{\bm{x}\in \square} \chi_{\bm{x}} \mid \chi\in B_{\mu}~| \equiv 1 \mod 2$; here $|C|$ is the cardinality of set $C$.
We finally get
\bea
    d \alpha_{\mu} (t) &=\frac{i}{\tr( B_\mu^2) }  \tr( B_{\mu} [dH(t), O(t)]   ) 
    - \sum_{\langle \bm{x} \bm{y}\rangle}\alpha_\mu (t) Vdt  - \alpha_\mu (t) Jdt\\
    &+\sum_{\langle \bm{x} \bm{y}\rangle}q_{\mu,\bm{x},\bm{y}}\alpha_\mu (t)Vdt
    +q_{\mu,\prod_{\bm{x}\in \square} \chi_{\bm{x}} }\alpha_\mu (t)Jdt\\
    &=\frac{i}{\tr( B_\mu^2) }  \tr( B_{\mu} [dH(t), O(t)]   )
    -2\sum_{\{\langle \bm{x} \bm{y}\rangle \mid q_{\mu,\bm{x},\bm{y}}=-1\}}\alpha_\mu (t) Vdt
    -2\delta_{ q_{\mu,\prod_{\bm{x}\in \square} \chi_{\bm{x}} },-1 }\alpha_\mu (t)Jdt.
\eea

Define $f_\mu(t )$ to be the average probability at time $t$
\bea
    f_\mu(t )  = \overline{|\alpha_{\mu}(t)|^2} 
    = \overline{\alpha^2_\mu(t)},
\eea
the evolution is given by
\bea
  d f_\mu(t ) = 2 \overline{\alpha_\mu(t) d \alpha_\mu(t) } + \overline{ d \alpha_\mu(t) d \alpha_\mu(t)  }.
\eea
We have
\bea
    d f_\mu(t) &= -4\sum_{\{\langle \bm{x} \bm{y}\rangle \mid q_{\mu,\bm{x},\bm{y}}=-1\}}f_\mu(t)Vdt
    -4\delta_{ q_{\mu,\prod_{\bm{x}\in \square} \chi_{\bm{x}} },-1 }f_\mu(t)Jdt
    -\frac{1}{\tr^2( B_\mu^2) }\tr^2( O(t) [B_{\mu},dH(t)]) \\
    &=-4\sum_{\{\langle \bm{x} \bm{y}\rangle \mid q_{\mu,\bm{x},\bm{y}}=-1\}}f_\mu(t)Vdt
    -4\delta_{ q_{\mu,\prod_{\bm{x}\in \square} \chi_{\bm{x}} },-1 }f_\mu(t)Jdt \\
    &+\sum_{\nu}\sum_{\langle \bm{x} \bm{y}\rangle}\frac{1}{\tr^2( B_\mu^2)} \tr^2(B_\nu [B_{\mu},\chi_{\bm{x}}\chi_{\bm{y}}]) f( B_\nu, t )Vdt 
    -\sum_{\nu}\frac{1}{\tr^2( B_\mu^2)} \tr^2(B_\nu [B_{\mu},\prod_{\bm{x}\in \square} \chi_{\bm{x}} ]) f( B_\nu, t )Jdt\\
    &=-4\sum_{\{\langle \bm{x} \bm{y}\rangle \mid q_{\mu,\bm{x},\bm{y}}=-1\}}f_\mu(t)Vdt
    -4\delta_{ q_{\mu,\prod_{\bm{x}\in \square} \chi_{\bm{x}} },-1 }f_\mu(t)Jdt \\
    &+4\sum_{\{\langle \bm{x} \bm{y}\rangle \mid q_{\mu,\bm{x},\bm{y}}=-1\},~\{\nu\big| |B_\mu\chi_{\bm{x}}\chi_{\bm{y}}|=|B_\nu|\}}f( B_\nu, t )Vdt
    +4\sum_{\{\nu\big| |B_\mu\prod_{\bm{x}\in \square} \chi_{\bm{x}}|=|B_\nu|\}}\delta_{ q_{\mu,\prod_{\bm{x}\in \square} \chi_{\bm{x}} },-1 }f( B_\nu, t )Jdt,
\eea
where the first equality uses the cyclicity of trace
\bea
    \tr( B_{\mu} [dH(t), O(t)]) = \tr( O(t) [B_{\mu},dH(t)]).
\eea

Now we can consider the operator height distribution function
\bea
    f(\mathbf{h},t) =|\alpha_\mu(t)|^2 \Big| _{\mathbf{h}_\mu=\mathbf{h}}.
\eea
which satisfies the master equation
\bea\label{eq:meq}
    \frac{\partial f( \mathbf{h}, t ) }{\partial t}
    &=-4\sum_{\langle \bm{x} \bm{y}\rangle}\delta_{h_{\bm{x}} \oplus h_{\bm{y}},1} V f( \mathbf{h}, t )
    -4\delta_{\sum_{\bm{x}\in \square}^{\oplus} h_{\bm{x}},1} J f( \mathbf{h}, t )
    +4\sum_{\langle \bm{x} \bm{y}\rangle}\delta_{h_{\bm{x}} \oplus h_{\bm{y}},1} V f(\mathbf{h}\oplus\mathbf{e}_{\bm{x}}\oplus\mathbf{e}_{\bm{y}}, t )
    +4\delta_{\sum_{\bm{x}\in \square}^{\oplus} h_{\bm{x}},1} J f(\mathbf{h}\oplus \sum_{\bm{x}\in \square}^{\oplus} \mathbf{e}_{\bm{x}}, t )\\
    &=4V\sum_{\langle \bm{x} \bm{y}\rangle}\delta_{h_{\bm{x}} \oplus h_{\bm{y}},1}
    \big(
    f(\mathbf{h}\oplus\mathbf{e}_{\bm{x}}\oplus\mathbf{e}_{\bm{y}}, t )-f( \mathbf{h}, t )
    \big)
    +4J\delta_{\sum_{\bm{x}\in \square}^{\oplus} h_{\bm{x}},1}
    \big(
    f(\mathbf{h}\oplus \sum_{\bm{x}\in \square}^{\oplus} \mathbf{e}_{\bm{x}}, t )-f( \mathbf{h}, t )
    \big),
\eea
where $\mathbf{e}_{\bm{x}}$ represents a vector that takes the value $1$ on site $\bm{x}$ and $0$ at all other sites. The sum $\oplus$ is taken modulo 2.

\section{The return probability of a random walk}

The return probability of a random walk is the cumulative probability up to time infinity that the walker comes back where it starts. In 1921, P\'olya~\cite{polya_uber_1921} proved that for a simple random walk, the return probability is $1$ when the dimension $d \le 2$, and strictly less than $1$ when $d \ge 3$. This section reviews the computation of the return probability for a (potentially long range) random walk in a $d$-dimensional cubic lattice using Fourier transform. 

To set the stage, let the walker start from the origin ${\bf 0}$ of the lattice $\mathbb{Z}^d$ and perform a (Markovian) random walk. For each time step, there is a probability distribution $f( {\bf x} )$ that determines the displacement $\vec{x}$. Define $q_n(\vec{x})$ to be the probability of the walker to reach position $\vec{x}$ {\it for the first time} at $t = n$. The probability the walker ever returns to the origin after $t = 0$ is 
\begin{equation}
    p_{\rm return} = \sum_{n=0}^{\infty} q_n( \vec{0} ). 
\end{equation}
For consistency, we require $q_n( \vec{0} ) = 0$. Slightly generalizing of the notion, the return probability of walker ever reaching position $\vec{x}$ is
\begin{equation}
    p_{\rm return}(\vec{x} ) = \sum_{n=0}^{\infty} q_n( \vec{x} ). 
\end{equation}

\begin{figure}[h]
\centering
\includegraphics[width=0.5\columnwidth]{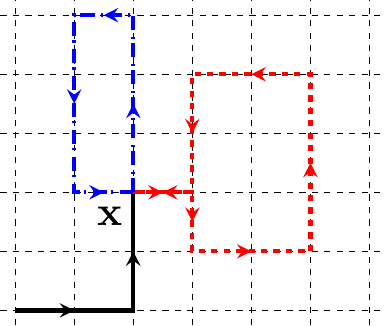}
\caption{A two-dimensional random walk starts from the origin at time $0$ and reaches position $\vec{x}$ at $t = 24$. The path is reducible that can be decomposed in three irreducible ones: the black path which contributes to $q_4(\vec{x})$ with $\vec{x}=(2,2)$, and the blue and red paths which contribute to $q_8({\bf 0})$ and $q_{12}({\bf 0})$ respectively.}
\label{fig:rw_path}
\end{figure}

\subsection{The recursion relation}
The probability $q_n( \vec{x} )$ is given by an irreducible path with the constraint of arriving at $\vec{x}$ for the first time. It is much easier to figure out the probability $p_n( \vec{x} )$ of unconstrained path that arrives at $\vec{x}$ at time $t$. Our initial condition is $p_0( \vec{x} ) = \delta_{ \vec{x}, \vec{0} }$. The probability $p_n(\vec{x})$ corresponds to reducible paths which can be decomposed into irreducible paths for each visit of $\vec{x}$, see Fig.~\ref{fig:rw_path}. This decomposition can be encapsulated into a recursion relation
\begin{equation}
\label{eq:recur_relation}
p_n ( \vec{x} ) = \sum_{k=0}^n p_{n-k}( \vec{0} )  q_k( \vec{x} )
\begin{cases}
    n \ge 1 & \vec{x} = \vec{0}\\
    n \ge 0 & \vec{x} \ne \vec{0} \\
\end{cases}.
\end{equation}
In words, the probability of arriving at $\vec{x}$ at time $t$ is the sum of probability of arriving at $\vec{x}$ for the first time at time $k$ and the probability of not moving for the rest $n-k$ steps. 

We can solve $q$ from $p$ through the generating functions $P(z, \vec{x}) =\sum_{n=0}^{\infty} p_n( \vec{x} ) z^n$ and $Q(z, \vec{x}) =\sum_{n=0}^{\infty} q_n( \vec{x} ) z^n$. The recursion relation in Eq.~\eqref{eq:recur_relation} translates to 
\begin{equation}
P( z, \vec{x} ) - \delta_{\vec{x}, \vec{0} } = P( z, \vec{0} ) Q ( z, \vec{x} ).
\end{equation}
Therefore
\begin{equation}
Q(z, \vec{x} ) 
=
\begin{cases}
    1 - \frac{1}{P( z, \vec{0} )}  & \vec{x} = \vec{0}\\
    \frac{P( z, \vec{x} ) }{P( z, \vec{0} )}  & \vec{x} \ne \vec{0} \\
\end{cases}.
\end{equation}

The return probability is
\begin{equation}
  p_{\rm return} (\vec{x} )  = Q( 1, \vec{x} ) 
   = \begin{cases}
    1 - \frac{1}{\sum_{n=0}^{\infty} p_n( \vec{0} ) } & \vec{x} = \vec{0}\\
    \frac{\sum_{n=0}^{\infty} p_n( \vec{x} ) }{ \sum_{n=0}^{\infty} p_n (\vec{0} ) } & \vec{x} \ne \vec{0} \\
\end{cases}.
\end{equation}

\subsection{Solution of the reducible probability}
For the random walk we consider, the probability satisfies a master equation: 
\begin{equation}
\label{eq:rw_master}
p_{n}(\vec{x}) = \sum_{{\vec{x}'}\in \mathbb{Z}^d} f(\vec{x} - \vec{x}') p_{n-1} (\vec{x}'). 
\end{equation}
The convolution becomes a product in Fourier space. Define the Fourier transform,
\begin{equation}
\begin{aligned}
    \tilde{p}_n(\vec{k} ) &= \sum_{\vec{x}\in \mathbb{Z}^d}  e^{ i \vec{k} \cdot \vec{x}} p_n( \vec{x} ),\\
p_n(\vec{x} ) &= \int_{[0,2\pi]^d} \prod_{j=1}^{d} \frac{dk_j}{2\pi} \tilde{p}_n(\vec{k}) e^{- i \vec{k} \cdot \vec{x} }.
\end{aligned}
\end{equation}
Then with the initial condition $\tilde{p}_0({\vec{k}}) = 1$, the master equation Eq.~\eqref{eq:rw_master} becomes
\begin{equation}
\tilde{p}_n (\vec{k}) = \tilde{f}({\vec{k}}) ^n \tilde{p}_0({\vec{k}}) = \tilde{f}(\vec{k})^n.
\end{equation}
Hence
\begin{equation}
  p_n(\vec{x}) = \int_{[0,2\pi]^d} \prod_{j=1}^{d} \frac{dk_j}{2\pi} \tilde{f}(\vec{k})^n e^{ -i \vec{k} \cdot \vec{x} }  
\end{equation}
and 
\begin{equation}
\label{eq:sum_p_n}
  \sum_{n=0}^{\infty} p_n(\vec{x}) = \int_{[0,2\pi]^d} \prod_{j=1}^{d} \frac{dk_j}{2\pi} \frac{1}{1 - \tilde{f}(\vec{k})}e^{ -i \vec{k} \cdot \vec{x} }. 
\end{equation}

\subsection{Simple random walks}

Let us specialize to simple random walk on $\mathbb{Z}^d$, which means the probability is isotropic in each lattice direction. The transition probability in Fourier space is
\begin{equation}
  f( \vec{k} ) = \frac{1}{d} \sum_{j=1}^d \cos ( k_j ). 
\end{equation}
For small $k$, the leading order expansion gives $f( \vec{k} ) \sim 1 - \frac{1}{2d} |\vec{k}|^2$. 

The return probability $p_{\rm return}( \vec{0} )$ is $1$ when $\sum_{n=0}^{\infty} p_n(\vec{0})$ diverges. From the power counting of Eq.~\eqref{eq:sum_p_n} around $k = |\vec{k}| \sim 0$ 
\begin{equation}
\sim \int_{[0,2\pi]^d} \prod_{j=1}^{d} \frac{dk_j}{2\pi} \frac{1}{k^2} 
\sim \int k^{d - 1} dk \frac{1}{k^2},
\end{equation}
we can see that the integral diverges for $d \le 2$ and is finite for $d > 2$ (meaning $d \ge 3$ for integer dimension), thus verifies P\'olya's theorem.

For $d = 3$, the return probability is
\begin{equation}
\begin{aligned}
      p_{\rm return} (\vec{0}) &=  1- \frac{1}{\int_{[0,2\pi]^3} \prod_{j=1}^{3} \frac{dk_j}{2\pi} \frac{1}{1- \frac{1}{3} ( \cos k_1 + \cos k_2 + \cos k_3 ) }} \\
      &\approx 0.340537.
\end{aligned}
\end{equation}

We can also analyze the scalings of $x = \vec{x}$ for $p_{\rm return}(\vec{x})$
\begin{equation}
\sim \int_{[0,2\pi]^d} \prod_{j=1}^{d} \frac{dk_j}{2\pi} \frac{1}{k^2}e^{ -i \vec{k} \cdot \vec{x} }  
\sim \int k^{d - 1} dk \frac{1}{k^2}  e^{ - ik x } \sim \frac{1}{x^{d-2}}.
\end{equation}

\subsection{L\'evy flight}

A $d$-dimensional L\'evy flight has a transition probability with long range tail
\begin{equation}
    f( \vec{x} ) \sim \frac{1}{|\vec{x}|^{d +\alpha} } \quad \alpha \in ( 0, 2 ]. 
\end{equation}
In Fourier space, 
\begin{equation}
\tilde{f}( \vec{k} ) \sim 1 - \# |\vec{k}|^{\alpha} 
\end{equation}
for small $k = |\vec{k}|$

The analysis of $p_{\rm return}( \vec{0} )$ and $p_{\rm return}( \vec{x} )$ can be carried over. 
\begin{equation}
p_{\rm return}( \vec{0} ) \sim \int k^{d - 1} dk \frac{1}{k^\alpha},
\end{equation}
which is convergent when $d > \alpha$. In other words, the return probability is $1$ when $\alpha > d$ and less than $1$  when $\alpha < d$. 

For the latter case, 
\begin{equation}
p_{\rm return}(\vec{x}) \sim \int k^{d - 1} dk \frac{1}{k^{\alpha}}  e^{ - ik x } \sim \frac{1}{x^{d-\alpha}}.
\end{equation}

\end{document}